\documentclass[12pt]{article}
\newcommand{\be}{\begin{equation}}
\newcommand{\ee}{\end{equation}}

\newcommand{\bi}{\begin{itemize}}
\newcommand{\ei}{\end{itemize}}
\newcommand{\bea}{\begin{eqnarray}}
\newcommand{\eea}{\end{eqnarray}}
\newcommand{\ba}{\begin{array}}
\newcommand{\ea}{\end{array}}

\usepackage{smartdiagram}
\usepackage[left=2.50cm, right=2.50cm, top=2.50cm, bottom=2.50cm]{geometry}
\usepackage[utf8]{inputenc}
\usepackage{cancel}
\usepackage{boldline}
\usepackage[english]{babel}
\usepackage{physics}
\usepackage{hyperref}
\usepackage{amsthm}
\usepackage{mathtools}
\usepackage{graphicx}
\usepackage{changepage}
\usepackage{amssymb,amsmath}
\usepackage{verbatim}
\usepackage{empheq}
\usepackage{xcolor}
\usepackage{tikz}
\usepackage{float}
\usepackage{multirow}
\usepackage{multicol}
\usepackage{amsmath}
\usepackage{centernot}
\usepackage[hang,flushmargin]{footmisc} %toglie indentazione da footnote
\usepackage[normalem]{ulem}
\usepackage{tabu}
\usetikzlibrary{tikzmark}
\usepackage{forest}
\numberwithin{equation}{section}
\hypersetup{hidelinks}
\newlength{\bibitemsep}
\newlength{\bibparskip}
\let\oldthebibliography\thebibliography
\renewcommand\thebibliography[1]{%
  \oldthebibliography{#1}%
  \setlength{\parskip}{\bibitemsep}%
  \setlength{\itemsep}{\bibparskip}%
}
\begin{document}
%%%%%%%%%%%%% Title %%%%%%%%%%%%%%%%%%%%%%%%%%
\par
\bigskip
\Large
\noindent
{\bf The theory of symmetric tensor field with boundary: Kac-Moody algebras in linearized gravity\\
}
%\bigskip
\par
\rm
\normalsize

%%%%%%%%%%%%%%%%%%%%%%%%%%%%%%%%%%%%%%%%%%%%%
%%%%%%%%%%%% Authors %%%%%%%%%%%%%%%%%%%%%%%%
\hrule

\vspace{1cm}

\large
\noindent
{\bf Erica Bertolini$^{1,2,a}$}, 
{\bf Nicola Maggiore$^{1,2,b}$}\\

\par
\small

\noindent$^1$ Dipartimento di Fisica, Universit\`a di Genova, via Dodecaneso 33, 16146 Genova, Italy.\\
\smallskip
\noindent$^2$ Istituto Nazionale di Fisica Nucleare (I.N.F.N.) - Sezione di Genova, via Dodecaneso 33, 16146 Genova, Italy.

\smallskip

\vspace{1cm}

\noindent
{\tt Abstract}
\\[20px]
In this paper we consider four dimensional (4D) linearized gravity (LG) with a planar boundary, where the most general boundary conditions are derived following Symanzik's approach. The boundary breaks diffeomorphism invariance and this results in  a breaking of the corresponding Ward identity. From this, on the boundary we find two conserved currents which form an algebraic structure of the Kac-Moody type, with a central charge proportional to the action ``coupling''. Moreover, we identify the boundary degrees of freedom, which are two symmetric rank-2 tensor fields, and derive the symmetry transformations, which are diffeomorphisms. The corresponding most general 3D action is obtained and a contact with the higher dimensional theory is established by requiring that the 3D equations of motion coincide with the 4D boundary conditions. Through this kind of holographic procedure, we find two solutions~: LG for a single tensor field and LG for two tensor fields with a mixing term. Curiously, we find that the Symanzik's 4D boundary term which governs the whole procedure contains a mass term of the Fierz-Pauli type for the bulk graviton.

\vspace{\fill}

\noindent{\tt Keywords:} \\
Quantum field theory with boundary, symmetric tensor gauge field theory, linearized gravity, boundary algebra.

\vspace{1cm}

\hrule
\noindent{\tt E-mail:
$^a$erica.bertolini@ge.infn.it,
$^b$nicola.maggiore@ge.infn.it.
}
\newpage

%*****************************************************
\section{Introduction}
%*****************************************************

In most cases, Quantum Field Theories (QFTs) are considered without boundaries and have been successful in providing descriptions of fundamental interactions, including gravity and cosmology. This is because one is generally interested in bulk effects, where the boundary can be neglected. Nevertheless, boundaries do exist, and in some cases, their effects are self-evident and dominant. The first example that comes to mind is the Casimir effect, which is the attractive force between two mirrors in a vacuum. This tiny effect, predicted in \cite{Casimir:1948dh}, was measured for the first time much later \cite{Lamoreaux1996}. This explicit boundary effect is due to the vacuum energy of QFT, and it is not surprising that it attracted Symanzik's interest in \cite{Symanzik:1981wd}, where, to our knowledge, the first rigorous description of a QFT with a boundary has been done. More recently, important phenomena pertaining to condensed matter physics, like the Fractional Quantum Hall Effect \cite{Stormer:1999zza} and the behavior of Topological Insulators \cite{Hasan:2010xy}, have been explained in terms of topological QFTs with boundaries \cite{Zee:1995avy,Wen:1995qn,Witten:2015aoa,Cho:2010rk,Savary:2016ksw,Kapustin:2014gua,Amoretti:2014iza,Bertolini:2021iku,Bertolini:2022sao}.
It is worth stressing that this is rather counterintuitive: topological QFTs, when considered without boundaries, have a vanishing Hamiltonian and no energy-momentum tensor. They might appear as the least physical theories one can imagine. Despite this, when a boundary is introduced, an extremely rich physics emerges, which can be observed experimentally. This is an example of the power of the boundary, which becomes even more striking in the gauge/gravity duality, or AdS/CFT correspondence \cite{Maldacena:2003nj}, particularly when this is, once again, applied in the context of condensed matter \cite{Hartnoll:2009sz}. In that case, the physics of strongly interacting systems, like superconductors or strange metals, is close to being understood by assuming that ordinary 4D spacetime is indeed a boundary of a 5D world dominated by gravity and populated by charged black holes of the Reissner-Nordstr\"om type.\\

In this paper, we consider four-dimensional (4D) Linearized Gravity (LG) with a planar boundary, motivated by the guessed relation between Kac-Moody (KM) algebras \cite{Kac:1967jr,Moody:1966gf} and LG \cite{Chamseddine:1988tu,Houart:2005wh,Hinterbichler:2022agn}. Indeed, KM algebras are tightly related to quantum field theory with a boundary. Starting from the pioneering paper \cite{Moore:1989yh}, where all rational 2D conformal field theories were derived from the 3D Chern-Simons theory with a boundary, it is now almost paradigmatic that conserved currents exist on the boundary of topological field theories, forming KM-type algebras with a central charge proportional to the inverse of the coupling constant of the bulk theory. These conserved currents are observables and observed indeed, at least in Chern-Simons and BF theories \cite{Birmingham:1991ty}, where they represent the chiral edge modes characterizing those phenomena. More recently, the same mechanism (boundary $\rightarrow$ conserved currents $\rightarrow$ KM algebras) has been reproduced in non-topological models, like Maxwell theory, for instance \cite{Blasi:2019wpq,Maggiore:2019wie,Bertolini:2020hgr}. Therefore, our aim is to verify the guessed relation between LG, which is not a topological theory, and KM algebras, providing a precise interpretation of the physical quantity characterizing KM algebras -- its central charge. According to the above general rule, the central charge should be related to the LG coupling constant, normally set to one by a redefinition of the field, which in LG is a symmetric rank-2 tensor field. Therefore the positivity of the central charge could also give us a way to determine the sign of the LG ``coupling'' constant, which in this case cannot be inferred from an energy constraint.
Apart from the KM algebra of conserved current, the introduction of a boundary yields more. The presence of a boundary in the 3D Chern-Simons theory induces a 2D theory, the Tomonaga-Luttinger theory of a chiral scalar field \cite{Tomonaga:1950zz,Luttinger:1963zz,Haldane:1981zza,Maggiore:2017vjf}, which we may refer to as a ``holographic'' reduction of the 3D bulk theory. Similar dimensional reductions -- lower dimensional theories induced by the presence of a boundary in higher-dimensional ones -- can be performed in other theories, both topological and non-topological \cite{Maggiore:2018bxr,Amoretti:2013xya,Bertolini:2023sqa}. 
We apply the same mechanism in the case of LG to find the holographically induced 3D theory of 4D LG with a planar boundary. As we will describe in detail in the paper, the holographic construction proceeds as follows. Current conservation is an equation that can be solved by expressing the 4D bulk fields in terms of 3D boundary degrees of freedom. We identify these degrees of freedom as the lower-dimensional fields of the holographic 3D theory. Moreover, we require that the most general transformation preserving the definition of the boundary fields is a symmetry of the induced 3D theory. This leads to an invariant 3D action. Completing this holographic correspondence, we require that the equations of motion (EoM) of the 3D theory coincide with the 4D boundary conditions (BC), by suitably tuning the parameters at our disposal. This is a non-trivial demand, and the existence of a solution is not immediately evident.\\

The paper is organized as follows. In Section 2, we consider the 4D LG theory with a planar boundary, implemented using a Heaviside theta distribution in the bulk action. We derive the most general BC following a method introduced by Symanzik, without imposing them by hand. The presence of the boundary breaks the invariance under diffeomorphisms, resulting in a breaking of the diffeomorphism Ward identity. From this, conserved currents and their KM algebra are derived using standard QFT methods. The central charge appears to be proportional to the inverse of the LG ``coupling'' constant, as in topological field theories. Section 3 focuses on the identification of the holographically induced 3D theory. By first solving the current conservation equation, we find the 3D degrees of freedom and determine the most general transformations that preserve their invariance. Remarkably, we discover that these transformations are diffeomorphisms, which is not obvious. Thus, diffeomorphism invariance emerges as a consequence of our procedure rather than a mere requirement. Once we have the quantum fields with their transformations, we arrive at the most general 3D action, satisfying the additional QFT requirements of locality and power counting. Using this, in Section 4 we establish the aforementioned holographic connection, yielding two solutions for the induced 3D theory. Our results are summarized in Section 5.

%*****************************************************
\section{The bulk}
%*****************************************************

%*****************************************************
\subsection{The model}
%*****************************************************

Linearized Gravity is the theory of a rank-2 symmetric tensor field $h_{\mu\nu}(x)$ on a flat Minkowskian background with signature $\mbox{diag}(-1,1,1,1)$. We introduce a planar boundary at $x^3=0$, adopting the following conventions concerning indices
	\begin{empheq}{align}
	\alpha,\beta,\gamma,...&=\{0,1,2,3\}\\
	a,b,c,...&=\{0,1,2\}\\
	\textsc{a,b,c},...&=\{1,2\}\ .
	\end{empheq}
Moreover, $x^\mu=(x^0,x^1,x^2,x^3)$ and $X^m=(x^0,x^1,x^2)$ are the bulk and boundary coordinates, respectively. The boundary is introduced in the LG action by means of a Heaviside step function, which confines the model on a half space with single-sided planar boundary
	\be\label{LG-Sbulk}
	S_{bulk} =
	\lambda \int d^4x\;\theta(x^3) \left(
	\frac{1}{4}F^\mu_{\ \mu\nu} F_\rho^{\ \rho\nu}-\frac{1}{6}F^{\mu\nu\rho}F_{\mu\nu\rho}
	\right)\ ,
	\ee
where $F^{\mu\nu\rho}(x)$ is the rank-3 tensor field strength associated to the tensor field $h_{\mu\nu}(x)$ introduced in \cite{Bertolini:2023sqa,Bertolini:2022ijb}
	\begin{equation}
	F_{\mu\nu\rho}=F_{\nu\mu\rho}=\partial_\mu h_{\nu\rho}+\partial_\nu h_{\mu\rho}-2\partial_\rho h_{\mu\nu}\ ,
	\label{Fmunurho}\end{equation}
satisfying the following properties
	\begin{empheq}{align}
	F_{\mu\nu\rho}+F_{\nu\rho\mu}+F_{\rho\mu\nu}&=0\\
	\epsilon_{\alpha\mu\nu\rho}\partial^{\mu}F^{\beta\nu\rho}&=0\ .
	\end{empheq}
The constant $\lambda$ in \eqref{LG-Sbulk} could be reabsorbed through a redefinition of $h_{\mu\nu}(x)$, however we maintain it in order to keep track of the bulk contributions. Due to the presence of the boundary at $x^3=0$ the $x^3$-derivative of the gauge field at $x^3=0$ must be considered as independent from $h_{\mu\nu}(x)$ \cite{Karabali:2015epa}, thus we define
	\be\label{ht}
	\tilde h_{\mu\nu}\equiv\partial_3h_{\mu\nu}|_{x^3=0}\ ,
	\ee
and the fields have the mass dimensions
	\be\label{dimh}
	[h_{\mu\nu}]=1\quad;\quad[\tilde h_{\mu\nu}]=2\ .
	\ee
The LG theory \eqref{LG-Sbulk} in the absence of a boundary is invariant under the infinitesimal diffeomorphism transformation
	\be\label{LG-diff}
	\delta_{diff} h_{\mu\nu}=\partial_\mu\Lambda_\nu+\partial_\nu\Lambda_\mu\ ,
	\ee
where $\Lambda(x)$ is a local vector parameter. LG is a gauge field theory, for which a gauge should be fixed. We choose the axial gauge, as customary in presence of a boundary \cite{Amoretti:2014iza}
	\be\label{h3mu=0}
	h_{\mu3}=0\ .
	\ee
This can be realized by means of a vector Lagrange multiplier $b^\mu(x)$ through the gauge-fixing term
	\be
	S_{gf}=\int d^4x\;\theta(x^3)b^\mu h_{\mu3}\ .
	\ee
Moreover, the following source term is needed
	\be
	S_J=\int d^4x\left[\theta(x^3)J^{ab}h_{ab}+\delta(x^3)\tilde J^{ab}\tilde h_{ab}\right]\ ,
	\ee
	 where, together with the external field $J^{ab}(x)$ associated to the tensor gauge field $h_{ab}(x)$, on the boundary an additional external source $\tilde J^{ab}(x)$ is coupled to $\tilde h_{ab}(x)$ \eqref{ht}. The presence of a boundary requires boundary conditions (BC). Instead of choosing a particular BC, we  follow Symanzik's approach \cite{Symanzik:1981wd}, where the most general BC for a QFT with boundary is achieved by adding to the action a boundary term constrained only by power-counting and locality. This modifies the EoM by a boundary contribution, and the BC are then obtained by means of a simple variational principle. For our model the boundary term is
	\be\label{LG-Sbd}
	S_{bd}=\int d^4x\delta(x^3)\left[\xi_0 h_{ab}h^{ab}+\xi_1\tilde h_{ab}h^{ab}+\xi_2\epsilon^{abc}h_{ai}\partial_bh_c^i+\xi_3h^2+\xi_4\tilde hh\right]\ ,
	\ee
where
	\be
	h\equiv\eta^{\mu\nu}h_{\mu\nu}\quad;\quad\tilde h\equiv\eta^{\mu\nu}\tilde h_{\mu\nu}\ ,
	\ee
and the $\xi_i,\ i=\{0,...,4\}$, are constant parameters with mass dimensions
	\be
	[\xi_0]=[\xi_3]=1\quad;\quad[\xi_1]=[\xi_2]=[\xi_4]=0\ .
	\ee
Notice that, due to the gauge-fixing condition \eqref{h3mu=0}, the nontrivial part of the trace $h(x)$ is $\eta^{ab}h_{ab}(x)$. The full action of the model finally is
	\be\label{Stot}
	S_{tot}=S_{bulk}+S_{gf}+S_{J}+S_{bd}\ .
	\ee

%*****************************************************
\subsection{Equations of motion and boundary conditions}
%*****************************************************

Besides the EoM of the Lagrange multiplier $b^\mu(x)$, which implements the axial gauge condition \eqref{h3mu=0}
	\be
	\frac{\delta S_{tot}}{\delta b^\mu}=h_{\mu3}=0\ ,
	\ee
the EoM of the gauge field $h_{\alpha\beta}(x)$ and its $\partial_3$-derivative $\tilde h_{\alpha\beta}(x)$ are, respectively
\be
\begin{split}
0=\frac {\delta S_{tot}}{\delta h_{\alpha\beta}} =&\theta(x^3)\left\{\lambda \left[-\partial_{\mu}F^{\alpha\beta\mu}
+\eta^{\alpha\beta}\partial_\mu F_\nu^{\ \nu\mu}-\tfrac{1}{2}\left(\partial^\alpha F_\mu^{\ \mu\beta}+\partial^\beta F_{\mu}^{\ \mu\alpha}\right)\right]+\delta^\alpha_a\delta^\beta_bJ^{ab}+\tfrac{1}{2}(b^\alpha\delta^\beta_3+b^\beta\delta^\alpha_3)\right\}+\\
&+\delta(x^3)\left\{\lambda \left[-F^{\alpha\beta3}+\eta^{\alpha\beta}F_\mu^{\ \mu3}-\tfrac{1}{2}\left(\eta^{\alpha3}F_\mu^{\ \mu\beta}+\eta^{\beta3}F_\mu^{\ \mu\alpha}\right)\right]+\right.\\
&+\left.\delta^\alpha_a\delta^\beta_b\left[ 2\xi_0h^{ab}+\xi_1\tilde h^{ab}+\xi_2(\epsilon^{aij}\partial_ih_j^b+\epsilon^{bij}\partial_ih_j^a)+2\xi_3\eta^{ab}h+\xi_4\eta^{ab}\tilde h\right]\right\}\ ,\label{LG-eomh}
\end{split}
\ee
and
\be
\begin{split}
0=\frac {\delta S_{tot}}{\delta \partial_3h_{\alpha\beta}} =&\lambda \theta(x^3)\left[F^{\alpha\beta3}
-\eta^{\alpha\beta}F_\mu^{\ \mu3}+\tfrac{1}{2}\left(\eta^{\alpha3}F_\mu^{\ \mu\beta}+\eta^{\beta3}F_{\mu}^{\ \mu\alpha}\right)\right]
+\delta(x^3)\delta^\alpha_a\delta^\beta_b\left\{ \tilde J^{ab}+\xi_1h^{ab}+\xi_4\eta^{ab}h\right\}\ .\label{LG-eomht}
\end{split}
\ee
The BC come from a variational principle applied on the EoM as $\lim_{\epsilon\to0}\int_0^\epsilon dx^3(\mbox{EoM})$, which corresponds to putting equal to zero the $\delta(x^3)$ contribution of the EoM \eqref{LG-eomh} and \eqref{LG-eomht}. From $\lim_{\epsilon\to0}\int^\epsilon_0dx^3\eqref{LG-eomh}$ we have
\be\label{LG-BC}
\begin{split}
&\left\{\lambda \left[-F^{\alpha\beta3}+\eta^{\alpha\beta}F_\mu^{\ \mu3}-\tfrac{1}{2}\left(\eta^{\alpha3}F_\mu^{\ \mu\beta}+\eta^{\beta3}F_\mu^{\ \mu\alpha}\right)\right]+\right.\\
&+\left.\delta^\alpha_a\delta^\beta_b\left[ 2\xi_0h^{ab}+\xi_1\tilde h^{ab}+\xi_2(\epsilon^{aij}\partial_ih_j^b+\epsilon^{bij}\partial_ih_j^a)+2\xi_3\eta^{ab}h+\xi_4\eta^{ab}\tilde h\right]\right\}_{x^3=0}=0\ .
\end{split}
\ee
The nontrivial components are
\begin{itemize}
\item $\alpha=3,\ \beta=b$ :
	\be
	\lambda \left(\partial^bh-\partial_ah^{ab}\right)_{x^3=0}=\lambda F_\mu^{\ \mu b}|_{x^3=0}=0\ .\label{LG-bch3i}
	\ee
\item $\alpha=a,\ \beta=b$ :
	\be
	\left[ 2\xi_0h^{ab}+(2\lambda +\xi_1)\tilde h^{ab}+\xi_2(\epsilon^{aij}\partial_ih_j^b+\epsilon^{bij}\partial_ih_j^a)+2\xi_3\eta^{ab}h+(\xi_4-2\lambda )\eta^{ab}\tilde h\right]_{x^3=0}=0\ .\label{LG-bchij}
	\ee
\end{itemize}
Taking $\lim_{\epsilon\to0}\int^\epsilon_0dx^3\eqref{LG-eomht}$ and going on-shell ($\tilde J=0$), we get 
\be
\delta^\alpha_a\delta^\beta_b\left(\xi_1h^{ab}+\xi_4\eta^{ab}h\right)_{x^3=0}=0\ ,\label{LG-bcht}
\ee
whose nonvanishing components are $\alpha=a,\ \beta=b$, which give
\be
\left( \xi_1h^{ab}+\xi_4\eta^{ab}h\right)_{x^3=0}=0\label{LG-bchtij}\ .
\ee
Notice that from \eqref{LG-bch3i} and taking $\partial_a$-derivative of \eqref{LG-bchtij} we have the following constraint on the boundary parameters
	\be
	\xi_1=-\xi_4\ .
	\ee
To summarize, the most general BC on the planar boundary $x^3=0$ are the following
\begin{empheq}{align}
\partial^bh-\partial_ah^{ab}&=0\label{BCh}\\
\xi_1\left( h^{ab}-\eta^{ab}h\right)&=0\label{bc2}\\
 2\xi_0h^{ab}+(2\lambda +\xi_1)&(\tilde h^{ab}-\eta^{ab}\tilde h)+\xi_2(\epsilon^{aij}\partial_ih_j^b+\epsilon^{bij}\partial_ih_j^a)+2\xi_3\eta^{ab}h=0\ .\label{bc3}
\end{empheq}
The BC \eqref{BCh} is universal, in the sense that it does not depend on $S_{bd}$ \eqref{LG-Sbd}. It represents the conservation of a current on the boundary
\be
\partial_a K^{ab}=0\ ,
\label{conscurrK}\ee
with
\be
K^{ab}\equiv h^{ab}-\eta^{ab}h\ .
\label{K}\ee
On the other hand, we remark that if $\xi_1=0$, the BC are given by \eqref{BCh} and \eqref{bc3}. If instead $\xi_1\neq0$, \eqref{bc2} implies \eqref{BCh} and the BC are given by \eqref{bc2} and \eqref{bc3}.

%*****************************************************
\subsection{Ward identities}
%*****************************************************

From the EoM for $h_{\mu\nu}(x)$ \eqref{LG-eomh} we get
	\be\label{LG-ward-ab}
	0=\int dx^3\partial_a\frac{\delta S_{tot}}{\delta h_{ab}}=2\lambda \left(\partial^b\tilde h-\partial_a\tilde h^{ab}\right)_{x^3=0}+\int dx^3\theta(x^3)\partial_aJ^{ab}\ ,
	\ee
where we used the BC \eqref{LG-BC}. We thus obtain the following Ward identity
	\be\label{LG-ward1}
	\int dx^3\theta(x^3)\partial_aJ^{ab}=-2\lambda \left(\partial^b\tilde h-\partial_a\tilde h^{ab}\right)_{x^3=0}\ ,
	\ee
 which is broken on the boundary $x^3=0$. 
 In the same way, from the EoM of $\tilde h_{\mu\nu}(x)$ \eqref{LG-eomht}, we find
	\be
		0=\int dx^3\partial_a\frac {\delta S_{tot}}{\delta \partial_3h_{ab}}=\partial_a\tilde J^{ab}|_{x^3=0}-2\lambda \left(-\partial_ah^{ab}+\partial^bh\right)_{x^3=0}\ ,
	\ee
which represents a local Ward identity, broken by the boundary
	\be\label{ward2}
	\partial_a\tilde J^{ab}|_{x^3=0}=2\lambda \left(-\partial_ah^{ab}+\partial^bh\right)_{x^3=0}\ .
	\ee
Notice that the r.h.s. describes the conservation on the boundary  of the current $K^{ab}(X)$ \eqref{K}, previously found as the BC \eqref{BCh}, hence we may write
	\be\label{ward2=0}
	\partial_a\tilde J^{ab}|_{x^3=0}=0\ .
	\ee
Going on-shell ($J=\tilde J=0$), the broken Ward identity \eqref{LG-ward1} yields
	\be
	\left(\partial^b\tilde h-\partial_a\tilde h^{ab}\right)_{x^3=0}=0\label{cc1}\ ,
	\ee
which, again, is a current conservation equation
\be
\partial_a \tilde K^{ab}=0\ ,
\label{conscurrKt}\ee
with
\be
\tilde{K}^{ab}\equiv \tilde{h}^{ab}-\eta^{ab}\tilde h\ .
\label{Kt}\ee
Hence, the presence of a planar boundary in LG theory has as a consequence the presence of conserved currents, which consist of the particular combinations \eqref{K} and \eqref{Kt}. This is remarkable because this is typical of topological field theories like 3D Chern-Simons and the BF models in any spacetime dimensions \cite{Birmingham:1991ty}.

%*****************************************************
\subsection{Kac-Moody Algebra}
%*****************************************************

By computing the functional derivative with respect to $J^{mn}(x')$ of the broken Ward identity \eqref{LG-ward1}, $i.e.$ $\frac{\delta}{\delta J^{mn}(x')}\eqref{LG-ward1}$ :
	\be
	\int_0^\infty dx^3\partial_a\left(\frac{\delta^a_m\delta^b_n+\delta^a_n\delta^b_m}{2}\delta^{(4)}(x-x')\right)=-2\lambda \left(\eta_{ac}\partial^b-\delta^b_c\partial_a\right)\frac{\delta Z_c[J,\tilde J]}{\delta\tilde J_{ac}\delta J'^{mn}}\ ,
	\ee
we get the commutation relations
	\be
	\frac{1}{2}\left(\delta^a_m\delta^b_n+\delta^a_n\delta^b_m\right)\partial_a\delta^{(3)}(X-X')
=-2i\lambda \left(\eta_{ac}\eta^{b0}-\delta^b_c\delta^0_a\right)\left[\tilde h^{ac}\ ,\ h'_{mn}\right]\delta(x^0-x'^0)\ ,
	\ee
where we used the on-shell constraint \eqref{cc1}. By setting
\bi
\item $b=0$ we have
	\be
	\left(\delta^a_m\delta^0_n+\delta^a_n\delta^0_m\right)\partial_a\delta^{(3)}(X-X')=4i\lambda \left[\tilde h^\textsc{d}_\textsc{d}\ ,\ h'_{mn}\right]\delta(x^0-x'^0)\ ,
	\ee
from which, integrating over time,
	\bi
	\item  $m=n=0$ gives
		\be
		\left[\tilde h^\textsc{d}_\textsc{d}\ ,\ h'_{00}\right]_{x^0=x'^0}=0\ .
		\ee
	\item $m=0,\ n=\textsc n$ we get
		\be
		\left[\tilde h^\textsc{d}_\textsc{d}\ ,\ h'_{0\textsc n}\right]_{x^0=x'^0}=-\frac{i}{4\lambda }\partial_\textsc{n}\delta^{(2)}(X-X')\ .
		\ee
This can be identified as a Kac-Moody (KM) algebraic structure \cite{Kac:1967jr,Moody:1966gf} with central charge
	\be\label{cc}
	c=-\frac{1}{4\lambda}\ ,
	\ee
which implies 
	\be
	\lambda<0\ ,
	\ee
because of the positivity of central charge of KM algebras \cite{mack,Becchi:1988nh}.
	\item $m=\textsc m,\ n=\textsc n$ gives
		\be
		\left[\tilde h^\textsc{d}_\textsc{d}\ ,\ h'_\textsc{mn}\right]_{x^0=x'^0}=0\ .
		\ee
	\ei
	\item $b=\textsc b$
	\be
	\left(\delta^a_m\delta^\textsc{b}_n+\delta^a_n\delta^\textsc{b}_m\right)\partial_a\delta^{(3)}(X-X')=4i\lambda \left[\tilde h^{0\textsc b}\ ,\ h'_{mn}\right]\delta(x^0-x'^0)\ .
	\ee
	\bi
	\item $m=n=0$
		\be
		\left[\tilde h^{0\textsc b}\ ,\ h'_{00}\right]_{x^0=x'^0}=0\ .
		\ee
	\item $m=0,\ n=\textsc n$
		\be
		\left[\tilde h^{0\textsc b}\ ,\ h'_{0\textsc n}\right]_{x^0=x'^0}=0\ .
		\ee
	\item $m=\textsc m,\ n=\textsc n$
		\be
		\left[\tilde h^{0\textsc b}\ ,\ h'_\textsc{mn}\right]_{x^0=x'^0}=-\frac{i}{4\lambda }\left(\delta^\textsc{a}_\textsc{m}\delta^\textsc{b}_\textsc{n}+\delta^\textsc{a}_\textsc{n}\delta^\textsc{b}_\textsc{m}\right)\partial_\textsc{a}\delta^{(2)}(X-X')\ .
		\ee
Here again a KM algebraic structure is observed with the same central charge  $c$ \eqref{cc}.
	\ei
\ei
We now compute the functional derivative of the broken Ward identity \eqref{LG-ward1} with respect to $\tilde J^{mn}(x')$, $i.e.\ \frac{\delta}{\delta \tilde J^{mn}(x')}\eqref{LG-ward1}$ :
	\be
	0=-2i\lambda \left(\eta_{ac}\eta^{b0}-\delta^b_c\delta^0_a\right)\left[\tilde h^{ac}\ ,\ \tilde h'_{mn}\right]_{x^0=x'^0}\ ,
	\ee
where we used the on-shell constraint \eqref{cc1} and integrated over time. In particular we have at
	\bi
	\item $b=0$
		\be
		\left[\tilde h^\textsc{d}_\textsc{d}\ ,\ \tilde h'_{mn}\right]_{x^0=x'^0}=0\ ;
		\ee
	\item $b=\textsc b$
		\be
		\left[\tilde h^{0\textsc b}\ ,\ \tilde h'_{mn}\right]_{x^0=x'^0}=0\ .
		\ee
	\ei
Summarizing, from the integrated Ward identity \eqref{LG-ward1} we get the semidirect sum of KM algebras with the same central charge 
	\begin{empheq}{align}
	\left[\tilde h^\textsc{d}_\textsc{d}\ ,\ h'_{0\textsc n}\right]&=-\frac{i}{4\lambda }\partial_\textsc{n}\delta^{(2)}(X-X')\label{[Trht,h0n]}\\
	\left[\tilde h^{0\textsc b}\ ,\ h'_\textsc{mn}\right]&=-\frac{i}{4\lambda }\left(\delta^\textsc{a}_\textsc{m}\delta^\textsc{b}_\textsc{n}+\delta^\textsc{a}_\textsc{n}\delta^\textsc{b}_\textsc{m}\right)\partial_\textsc{a}\delta^{(2)}(X-X')\label{[ht0b,hmn]}\ .
	\end{empheq}
The above KM algebraic structure has a physical meaning when expressed in terms of the conserved currents $K^{ab}(X)$ \eqref{K} and $\tilde K^{ab}(X)$ \eqref{Kt}, which are expressed in terms of the tensor fields $h^{ab}(X)$, $\tilde h^{ab}(X)$ and their traces. In fact, as a consequence of \eqref{[Trht,h0n]} and \eqref{[ht0b,hmn]} we find that $K^{ab}(X)$ and $\tilde K^{ab}(X)$ form a KM algebra with central charge \eqref{cc} whose non vanishing components are
	\begin{empheq}{align}
	\left[\tilde K^{00}\ ,\ K'_{0\textsc m}\right]&=-\frac{i}{4\lambda }\partial_\textsc{m}\delta^{(2)}(X-X')\label{KK1}\\
	\left[\tilde K^{0\textsc b}\ ,\ K'_{mn}\right]&=-\frac{i}{4\lambda }\left(\delta^\textsc{a}_m\delta^\textsc{b}_n+\delta^\textsc{a}_n\delta^\textsc{b}_m-2\eta^{\textsc a\textsc b}\eta_{mn}\right)\partial_\textsc{a}\delta^{(2)}(X-X')\label{KK2}\ .
	\end{empheq}
The existence of a KM algebraic structure for conserved currents on the boundary of 4D LG confirms the guess made in \cite{Hinterbichler:2022agn} as a particularly interesting possibility in connection with Weinberg's soft graviton theorems \cite{steve,He:2014laa,Kapec:2015vwa}.
	
%*****************************************************
\section{The boundary}
%*****************************************************

%*****************************************************
\subsection{The degrees of freedom}
%*****************************************************

The presence of a 3D boundary in the 4D theory described by the action $S_{tot}$ \eqref{Stot} induces a 3D theory, whose field content is determined by the solution of the on-shell broken Ward identity \eqref{LG-ward1}
	\be
	\partial_a\left(\tilde h^{ab}-\eta^{ab}\tilde h\right)_{x^3=0}=0\label{cc1'}
	\ee
and of the BC \eqref{BCh}
	\be
	\partial_a\left(h^{ab}-\eta^{ab}h\right)_{x^3=0}=0\ .\label{cc2'}
	\ee
Let us consider first \eqref{cc1'}. Define 
	\be
	\tilde C^{ab}\equiv\tilde h^{ab}-\eta^{ab}\tilde h\ ,
	\ee
whose trace is
	\be
	\tilde C=\eta_{ab}\tilde C^{ab}=-2\tilde h\ .
	\ee
Eq. \eqref{cc1'} then reads
	\be\label{divC=0}
	\partial_a\tilde C^{ab}=0\ .
	\ee
In order to find the most general solution, let us parametrize the symmetric tensor $\tilde C^{ab}(X)$ as follows
	\be\label{solC1}
	\tilde C^{ab}=\frac{1}{2}\left(\epsilon^{amn}\partial_m\tilde \Sigma_n^{\ b}+\epsilon^{bmn}\partial_m\tilde \Sigma_n^{\ a}\right)\ .
	\ee
Because of \eqref{divC=0} it must be
	\be
	\epsilon^{bmn}\partial_m\partial_a\tilde \Sigma_n^{\ a}=0\ ,
	\ee
which is solved by
\be
\tilde \Sigma_n^{\ a}=\epsilon^{acd}\partial_c\tilde\sigma_{nd}+\partial_n\phi^a\ ,
\ee
but we observe that the $\phi^a(X)$ contribution trivializes $\tilde C^{ab}(X)$ \eqref{solC1}. Hence
	\be
	\tilde\Sigma_n^{\ a}=\epsilon^{acd}\partial_c\tilde\sigma_{nd}\ .
	\ee
In terms of this result, $\tilde C^{ab}(X)$ \eqref{solC1} solves \eqref{divC=0}, and reads
	\be\label{solC}
	\tilde C^{ab}=\epsilon^{bmn}\epsilon^{acd}\partial_m\partial_c\tilde\sigma_{nd}\ ,
	\ee
with $\tilde\sigma_{ab}(X)=\tilde\sigma_{ba}(X)$ as a consequence of the fact that $\tilde C^{ab}(X)$ is symmetric $\tilde C^{ab}(X)=\tilde C^{ba}(X)$, and with $[\tilde\sigma]=0$. Thus the general solution for $\tilde h^{ab}(X)$ is
	\be\label{h solved}
		\begin{split}
		\tilde h^{ab}&=\tilde C^{ab}-\frac{1}{2}\eta^{ab}\tilde C\\
		&=-\frac{1}{2}\eta^{ab}(\partial_m\partial^m\tilde\sigma_{n}^{ \;n}-\partial^m\partial^n\tilde\sigma_{mn})+\partial_m\partial^m\tilde\sigma^{ab}+\partial^a\partial^b\tilde\sigma_{n}^{ \;n}-\partial_c(\partial^b\tilde\sigma^{am}+\partial^a\tilde\sigma^{bc})\ .
		\end{split}
	\ee
The Eq.\eqref{cc2'} for $h^{ab}(x)$ has the same structure as \eqref{cc1'}, therefore the solution has the same form \eqref{h solved}. We finally get	
	\begin{empheq}{align}
		\tilde h^{ab}&=\epsilon^{bmn}\epsilon^{acd}\partial_m\partial_c\tilde\sigma_{nd}+\frac{1}{2}\eta^{ab}(\partial_m\partial^m\tilde\sigma_{n}^{\;n}-\partial^m\partial^n\tilde\sigma_{mn})\label{solht}\\
		&=-\frac{1}{2}\eta^{ab}(\partial_m\partial^m\tilde\sigma_{n}^{\;n}-\partial^m\partial^n\tilde\sigma_{mn})+\partial_m\partial^m\tilde\sigma^{ab}+\partial^a\partial^b\tilde\sigma_{n}^{\;n}-\partial_c(\partial^b\tilde\sigma^{ac}+\partial^a\tilde\sigma^{bc})\nonumber\\
		h^{ab}&=\epsilon^{bmn}\epsilon^{acd}\partial_m\partial_c \sigma_{nd}+\frac{1}{2}\eta^{ab}(\partial_m\partial^m \sigma_{n}^{\;n}-\partial^m\partial^n \sigma_{mn})\label{solh}\\
		&=-\frac{1}{2}\eta^{ab}(\partial_m\partial^m \sigma_{n}^{\;n}-\partial^m\partial^n \sigma_{mn})+\partial_m\partial^m \sigma^{ab}+\partial^a\partial^b \sigma_{n}^{\;n}-\partial_c(\partial^b \sigma^{ac}+\partial^a \sigma^{bc})\ ,\nonumber
	\end{empheq}
which means that the fields of the induced 3D theory are identified as the rank-2 symmetric tensors $\sigma^{ab}(X)$ and $\tilde\sigma^{ab}(X)$. Moreover, these solutions are invariant under the following transformations of the boundary fields $\sigma_{ab}(X),\ \tilde\sigma_{ab}(X)$
	\begin{empheq}{align}
	\tilde\delta\tilde h_{ab}=0\quad&\Leftrightarrow\quad\tilde\delta\tilde\sigma_{mn}=\partial_m\tilde\xi_n+\partial_n\tilde\xi_m\label{tdiff}\\
	\delta h_{ab}=0\quad&\Leftrightarrow\quad\delta\sigma_{mn}=\partial_m\xi_n+\partial_n\xi_m\ ,\label{diff}
	\end{empheq}
which remarkably means that the induced boundary theory must be invariant under infinitesimal diffeomorphisms, which therefore is a consequence of the general method we followed to introduce a boundary in LG, without need of requiring it explicitly.

%*****************************************************
\subsection{Most general 3D action}
%*****************************************************

As a consequence of the solutions $\tilde h_{ab}(x)$ \eqref{solht} and $h_{ab}(x)$ \eqref{solh} and of their mass dimensions \eqref{dimh}, the boundary fields $\sigma_{ab}(X)$ and $\tilde\sigma_{ab}(X)$ should have mass dimensions $[\sigma]=-1$ and $[\tilde\sigma]=0$. However, in 3D the canonical choices for the mass dimensions of the tensor fields are two :
	\begin{enumerate}
	\item $[\sigma]=[\tilde\sigma]=1$, which can be realized by rescaling as follows
		\be
		\tilde\sigma\to \tilde M^{-1}\tilde\sigma\quad;\quad\sigma\to M^{-2}\sigma\ .
		\ee
However in this case power-counting and locality constrain the action to the following Chern-Simons/BF-like action \cite{Birmingham:1991ty}, 
	\be\label{Scs}
	S=\int d^3x\epsilon^{abc}\left(a_1\sigma_{ad}\partial_b\sigma_c^{\ d}+a_2\tilde\sigma_{ad}\partial_b\sigma_c^{\ d}+a_3\tilde\sigma_{ad}\partial_b\tilde\sigma_c^{\ d}\right)
	\ee
which is not invariant under the diffeomorphism transformations $\delta$ \eqref{diff} and $\tilde\delta$ \eqref{tdiff} :
	\begin{empheq}{align}
	\delta S&=\int d^3x\epsilon^{abc}\left(2a_1\sigma_{ad}+a_2\tilde\sigma_{ad}\right)\partial_b\partial^d\xi_c\\
	\tilde\delta S&=\int d^3x\epsilon^{abc}\left(a_2\sigma_{ad}+2a_3\tilde\sigma_{ad}\right)\partial_b\partial^d\tilde\xi_c\ ,
	\end{empheq}
which indeed vanish only at the trivial case ($a_1=a_2=a_3=0$). Thus we must discard this possibility.
	\item $[\sigma]=[\tilde\sigma]=\frac{1}{2}$, achieved by rescaling
	\be\label{rescaling}
	\tilde\sigma\to \tilde M^{-\frac{1}{2}}\tilde\sigma\quad;\quad\sigma\to M^{-\frac{3}{2}}\sigma\ ,
	\ee
which, instead, leads to a nontrivial solution, as we shall see in what follows.
	\end{enumerate}
The most general action invariant under the infinitesimal diffeomorphisms $\tilde\delta$ \eqref{tdiff} and $\delta$ \eqref{diff} has the following structure
	\be\label{Sinv}
	S_{3D}[\sigma,\tilde\sigma]=\kappa S_{LG}[\sigma]+\tilde\kappa\tilde S_{LG}[\tilde\sigma]+\kappa_{m}S_{mix}[\sigma,\tilde\sigma]\ ,
	\ee
where $\kappa,\ \tilde\kappa,\ \kappa_m$ are dimensionless constants, and $S_{LG}[\sigma]$ and $\tilde S_{LG}[\tilde\sigma]$ are LG contributions analogous to \eqref{LG-Sbulk}, written in terms of the boundary tensor field $\sigma_{ab}(X)$ and $\tilde\sigma_{ab}(X)$, respectively
	\begin{empheq}{align}
	S_{LG} &= \int d^3x \left(\frac{1}{4}f^a_{\ ac} f_b^{\ bc}-\frac{1}{6}f^{abc}f_{abc}\right)\\
	\tilde S_{LG}&= \int d^3x \left(\frac{1}{4}\tilde f^a_{\ ac}\tilde f_b^{\ bc}-\frac{1}{6}\tilde f^{abc}\tilde f_{abc}\right)\ ,
	\end{empheq}
with
	\begin{empheq}{align}
	f_{abc}&=f_{bac}=\partial_a \sigma_{bc}+\partial_b \sigma_{ac}-2\partial_c \sigma_{ab}	\label{fabc}\\
	\tilde f_{abc}&=\tilde f_{bac}=\partial_a \tilde\sigma_{bc}+\partial_b \tilde\sigma_{ac}-2\partial_c \tilde\sigma_{ab}\ ,
	\label{tfabc}
	\end{empheq}
satisfying the ciclicity property
	\begin{empheq}{align}\label{cicl f}
	f^{abc}+f^{bca}+f^{cab}&=0\\
	 \tilde f^{abc}+\tilde f^{bca}+\tilde f^{cab}&=0\ .\label{cicl tf}
	\end{empheq}
Notice that no Chern-Simons or BF contributions like in \eqref{Scs} are allowed as a consequence of the diffeomorphism invariances $\delta S_{3D} =\tilde\delta S_{3D} =0$. The $S_{mix} $ term in \eqref{Sinv} is the most general one depending on both $\sigma_{ab}(X)$ and $\tilde\sigma_{ab}(X)$, compatible with power-counting and the invariances $\delta S_{mix}=\tilde \delta S_{mix}=0$. Excluding again Chern-Simons/BF-like contributions, which are not invariant under diffeomorphisms, we have
	\begin{equation}
		S_{mix} =\int d^3x\left\{a_0\partial_a\sigma\partial^a\tilde\sigma+a_1\partial_c\sigma_{ab}\partial^c\tilde\sigma^{ab}+a_2\partial_a\sigma\partial_b\tilde\sigma^{ab}+a_3\partial_a\tilde\sigma\partial_b\sigma^{ab}+a_4\partial_c\sigma_{ab}\partial^a\tilde\sigma^{bc}\right\}\ .
	\end{equation}
Imposing invariance under $\delta$ \eqref{diff} we get
	\begin{equation}
		\begin{split}
		\delta S_{mix} =0
		&=-\int d^3x\left\{\tilde\sigma^{ab}\left[\left(2a_1+a_4\right)\partial_a\partial^2\xi_b+\left(2a_2+a_4\right)\partial_a\partial_b\partial_m\xi^m\right]+
2\tilde\sigma\left(a_0+a_3\right)\partial_m\partial^2\xi^m\right\}\ ,
		\end{split}
	\end{equation}
which gives
	\be
		\begin{split}
		&a_3=-a_0\\
		&a_4=-2a_1\\
		&a_2=a_1	\ .
		\end{split}
	\ee
The $\delta$-invariant $S_{mix} $ action term is
	\begin{equation}\label{Smix}
		S_{mix} =\int d^3x\left\{a_0\partial_a\sigma\partial^a\tilde\sigma+a_1\partial_c\sigma_{ab}\partial^c\tilde\sigma^{ab}+a_1\partial_a\sigma\partial_b\tilde\sigma^{ab}-a_0\partial_a\tilde\sigma\partial_b\sigma^{ab}-2a_1\partial_c\sigma_{ab}\partial^a\tilde\sigma^{bc}\right\}\ .
	\end{equation}
Requiring now invariance under $\tilde\delta$ \eqref{tdiff}, we get
\begin{equation}
		\begin{split}
		\tilde\delta S_{mix} =0
		&=2\int d^3x\left\{\sigma_{ab}\left(a_0+a_1\right)\partial^a\partial^b\partial^m\tilde\xi_m
-\sigma\left(a_0+a_1\right)\partial_m\partial^2\tilde\xi^m\right\}\ ,
		\end{split}
	\end{equation}
hence it must be
	\be
	a_1=-a_0\ .
	\ee
Therefore the mixed action term \eqref{Smix} invariant under both $\tilde\delta$ \eqref{tdiff} and $\delta$ \eqref{diff} is
	\begin{equation}\label{Smix'}
		S_{mix} =a_0\int d^3x\left\{\partial_a\sigma\partial^a\tilde\sigma-\partial_c\sigma_{ab}\partial^c\tilde\sigma^{ab}-\partial_a\sigma\partial_b\tilde\sigma^{ab}-\partial_a\tilde\sigma\partial_b\sigma^{ab}+2\partial_c\sigma_{ab}\partial^a\tilde\sigma^{bc}\right\}\ .
	\end{equation}
After reabsorbing the $a_0$ parameter into $\kappa_m$ in \eqref{Sinv}, we observe that using the definitions of $f^{abc}(X)$ \eqref{fabc} and $\tilde f^{abc}(X)$ \eqref{tfabc} $S_{mix} $ \eqref{Smix'} can be written as
	\be\label{Smix''}
	S_{mix} =\int d^3x \left(\frac{1}{4} f^a_{\ ac}\tilde f_b^{\ bc}-\frac{1}{6} f^{abc}\tilde f_{abc}\right)\ .
	\ee
The most general invariant action therefore is
	\be\label{Sinv'}
		\begin{split}
		S_{3D} &=\kappa S_{LG} +\tilde\kappa \tilde S_{LG}+\kappa_{m}S_{mix} \\
		&=\int d^3x\left\{\kappa\left(\tfrac{1}{4}f^a_{\ ac} f_b^{\ bc}-\tfrac{1}{6}f^{abc}f_{abc}\right)+\tilde\kappa\left(\tfrac{1}{4}\tilde f^a_{\ ac}\tilde f_b^{\ bc}-\tfrac{1}{6}\tilde f^{abc}\tilde f_{abc}\right)+\kappa_m\left(\tfrac{1}{4} f^a_{\ ac}\tilde f_b^{\ bc}-\tfrac{1}{6} f^{abc}\tilde f_{abc}\right)\right\}\ .
		\end{split}
	\ee
We finally observe that $S_{mix} $ \eqref{Smix''} can be written as
	\be
	S_{mix} =\int d^3x\epsilon^{abc}\epsilon^{def}\sigma_{ad}\partial_b\partial_e\tilde\sigma_{cf}\ ,
	\ee 
hence, replacing $\tilde\sigma_{ab}(X)$ with $\sigma_{ab}(X)$ we have an alternative way to write the 3D LG action
	\be
	\begin{split}
	S_{LG} 
	&=\int d^3x(\epsilon^{abc}\partial_b\sigma_{am})(\epsilon^{pnm}\partial_n\sigma_{pc})\ ,
	\end{split}
	\ee 
whose EoM are
	\be
	\epsilon^{ap_1p_2}\epsilon^{bp_3p_4}\partial_{p_1}\partial_{p_3}h_{p_2p_4}=0\ ,
	\ee
which are those of LG written in an alternative and more compact way. A similar expression holds for 4D LG, whose EoM can be written as
	\be
	\epsilon^{\mu\alpha_1\alpha_2\alpha_3}\epsilon^{\nu\alpha_4\alpha_5\alpha_6}\eta_{\alpha_3\alpha_6}\partial_{\alpha_1}\partial_{\alpha_4}h_{\alpha_2\alpha_5}=0\ .
	\ee

%*****************************************************
\subsection{Equations of motion of the 3D induced theory}
%*****************************************************

From
	\begin{empheq}{align}
	\frac{\delta f_{abc}(x)}{\delta \sigma_{mn}(y)}&=[-(\delta^m_a\delta^n_b+\delta^m_b\delta^n_a)\partial_c+\tfrac{1}{2}(\delta^m_c\delta^n_b+\delta^m_b\delta^n_c)\partial_a+\tfrac{1}{2}(\delta^m_a\delta^n_c+\delta^m_c\delta^n_a)\partial_b]\delta^{(3)}(x-y)\\
	\frac{\delta f^a_{\ ac}(x)}{\delta \sigma_{mn}(y)}&=\left(-2\eta^{mn}\partial_c+\delta^n_c\partial^m+\delta^m_c\partial^n\right)\delta^{(3)}(x-y)\ ,
	\end{empheq}
and the ciclicity property of $f^{abc}(X)$ and $\tilde f^{abc}(X)$ \eqref{cicl f}, \eqref{cicl tf}, we find the following EoM for the boundary fields $\sigma_{ab}(X)$ and $\tilde\sigma_{ab}(X)$
	\begin{empheq}{align}
		\frac{\delta S_{3D} }{\delta\sigma_{mn}}=&\kappa\left[-\partial_af^{mna}+\eta^{mn}\partial_af_b^{\ ba}-\tfrac{1}{2}\left(\partial^mf_b^{\ bn}+\partial^nf_b^{\ bm}\right)\right]+\label{eomSig}\\
&+\frac{\kappa_m}{2}\left[-\partial_a \tilde f^{mna}+\eta^{mn}\partial_a\tilde f_b^{\ ba}-\tfrac{1}{2}\left(\partial^m\tilde f_b^{\ bn}+\partial^n\tilde f_b^{\ bm}\right)\right]=0\nonumber\\[5px]
		\frac{\delta S_{3D} }{\delta\tilde\sigma_{mn}}=&\tilde\kappa\left[-\partial_a\tilde f^{mna}+\eta^{mn}\partial_a\tilde f_b^{\ ba}-\tfrac{1}{2}\left(\partial^m\tilde f_b^{\ bn}+\partial^n\tilde f_b^{\ bm}\right)\right]+\label{eomtSig}\\
&+\frac{\kappa_m}{2}\left[-\partial_a  f^{mna}+\eta^{mn}\partial_a f_b^{\ ba}-\tfrac{1}{2}\left(\partial^m f_b^{\ bn}+\partial^n f_b^{\ bm}\right)\right]=0\ .\nonumber
	\end{empheq}	

%*****************************************************
\section{Contact between bulk and boundary}
%*****************************************************

It is possible to make a holographic contact between the 4D bulk theory described by the action $S_{tot}$ \eqref{Stot} and the induced 3D theory whose action is $S_{3D} $ \eqref{Sinv'} by requiring that the EoM \eqref{eomSig} and \eqref{eomtSig} derived from $S_{3D} $ coincide with the BC \eqref{BCh}, \eqref{bc2} and \eqref{bc3} we found for the 4D bulk theory. This can be achieved by suitably fine tuning the $\xi_i$ parameters appearing in $S_{bd}$ \eqref{LG-Sbd}, and $\kappa,\ \tilde\kappa,\ \kappa_m$ in $S_{3D} $ \eqref{Sinv'}.
The first step is to write the BC \eqref{BCh}, \eqref{bc2} and \eqref{bc3} in terms of the boundary fields $\sigma_{ab}(X)$ and $\tilde\sigma_{ab}(X)$ through the solutions \eqref{solht} and \eqref{solh}. The BC \eqref{BCh} is the defining equation for $h_{ab}(X)$ on the boundary \eqref{solh}, thus  the contact is automatically satisfied. Concerning  \eqref{bc2}, using \eqref{solh} we have, on $x^3=0$
	\be
	0= h^{ab}-\eta^{ab}h=M^{-\frac{3}{2}}\left[\partial^2\sigma^{ab}+\partial^a\partial^b\sigma-\partial_c\left(\partial^a\sigma^{bc}+\partial^b\sigma^{ac}\right)+\eta^{ab}\left(\partial^c\partial^d\sigma_{cd}-\partial^2\sigma\right)\right]
	\ee
where $M$ is the rescaling factor of $\sigma_{ab}(X)$ introduced in \eqref{rescaling}. This can also be written as
	\be
	H^{mn}\equiv\left( h^{mn}-\eta^{mn}h\right)|_\eqref{solh}=\frac{M^{-\frac{3}{2}}}{2}\left[-\partial_af^{mna}+\eta^{mn}\partial_af_b^{\ ba}-\tfrac{1}{2}\left(\partial^mf_b^{\ bn}+\partial^nf_b^{\ bm}\right)\right]\ .\label{h-f}
	\ee
Analogously
	\be
	\tilde H^{mn}\equiv\left(\tilde h^{mn}-\eta^{mn}\tilde h\right)|_\eqref{solh}=\frac{\tilde M^{-\frac{1}{2}}}{2}\left[-\partial_a\tilde f^{mna}+\eta^{mn}\partial_a\tilde f_b^{\ ba}-\tfrac{1}{2}\left(\partial^m\tilde f_b^{\ bn}+\partial^n\tilde f_b^{\ bm}\right)\right]\ ,\label{th-tf}
	\ee
where $\tilde M$ is the rescaling factor of $\tilde\sigma_{ab}(X)$ introduced in \eqref{rescaling}. We introduced $H^{ab}$ and $\tilde H^{ab}$ so that the contact between the bulk BC and the boundary EoM will be more evident, as we shall see. Indeed the 3D EoM \eqref{eomSig} and \eqref{eomtSig} can be written as linear combination of \eqref{h-f} and \eqref{th-tf}
	\be\label{H+tH}
	\alpha H^{mn}+\beta \tilde H^{mn}=0\ .
	\ee
Explicitly we have
	\begin{empheq}{align}
	\eqref{eomSig}\ =\ 2{\kappa}M^{\frac{3}{2}} H^{ab}+\kappa_m \tilde M^{\frac{1}{2}}\tilde H^{ab}&=0\label{eomH1''}\\
	\eqref{eomtSig}\ =\ \kappa_mM^{\frac{3}{2}}H^{ab}+2\tilde\kappa\tilde M^{\frac{1}{2}}\tilde H^{ab}&=0\label{eomH2''}\ .
	\end{empheq}
The BC \eqref{bc2} can be written as
	\be\label{bc2H}
	\xi_1H^{ab}=0\ ,
	\ee
while the BC \eqref{bc3} cannot be written as \eqref{H+tH}
	\be\label{BCHh}
	 2\xi_0H^{ab}+(2\lambda +\xi_1)\tilde H^{ab}+\xi_2(\epsilon^{aij}\partial_ih_j^b+\epsilon^{bij}\partial_ih_j^a)+2(\xi_3+\xi_0)\eta^{ab}h=0\ ,
	\ee
unless
	\be
	\xi_3=-\xi_0\quad;\quad \xi_2=0\ ,
	\ee
in which case the BC \eqref{bc3} becomes
	\be
	2\xi_0H^{ab}+(2\lambda +\xi_1)\tilde H^{ab}=0\ ,\label{bc2''}
	\ee
recalling that $\lambda$ is the coefficient of the bulk action $S_{bulk}$ \eqref{LG-Sbulk}. Now that both EoM and BC have a similar structure, we can $holographycally$ match them by tuning their parameters so that (EoM)$\leftrightarrow$(BC). Keeping in mind that the $\xi_1$ parameter defines two situations
	\bi
	\item $\xi_1=0$ : one BC \eqref{bc2''}
	\item $\xi_1\neq0$ : two BC \eqref{bc2H} and \eqref{bc2''},
	\ei
let us look at the first case.
	\begin{enumerate}
	\item $\pmb{\xi_1=0}$ : the only BC is, after a multiplication by $\frac{1}{2\lambda}$ (remember that $\lambda\neq0$, being the coupling constant of the bulk)
	\be\label{bc1H}
	\frac{\xi_0}{\lambda}H^{ab}+\tilde H^{ab}=0\ .
	\ee
In the same way we have seen that the EoM \eqref{eomSig} and \eqref{eomtSig} can be written as \eqref{eomH1''}, \eqref{eomH2''}
	\begin{empheq}{align}
	2\mu\frac{\kappa}{\kappa_m} H^{ab}+\tilde H^{ab}&=0\label{eomH1}\\
	\frac{\mu}{2}\frac{\kappa_m}{\tilde\kappa}H^{ab}+\tilde H^{ab}&=0\label{eomH2}\ ,
	\end{empheq}
where $\mu\equiv\sqrt{\tfrac{M^3}{\tilde M}}$ with $[\mu]=1$. They both match with the BC \eqref{bc1H} if
	\be\label{hc xi1=0}
	\frac{\xi_0}{\lambda}=2\mu\frac{\kappa}{\kappa_m}=\frac{\mu}{2}\frac{\kappa_m}{\tilde\kappa}\quad\Rightarrow\quad\kappa_m^2=4\kappa\tilde\kappa\ ,\ \kappa\tilde\kappa>0\ .
	\ee
The implication on the 3D action \eqref{Sinv'} is that the following redefinition of the fields is possible 
	\be\label{rho}
	\rho_{ab}\equiv\sqrt\kappa\sigma_{ab}\pm\sqrt{\tilde\kappa}\tilde\sigma_{ab}\quad;\quad\Phi_{abc}\equiv\sqrt\kappa f_{abc}\pm\sqrt{\tilde\kappa}\tilde f_{abc}\ ,
	\ee
such that the action only depends on one field as 
	\be
		\begin{split}
		S_{3D}&=\frac{1}{6}\int d^3x\left(\frac{1}{\sqrt2}\eta^{ab}\Phi_m^{\ mc}-\Phi^{abc}\right)\left(\frac{1}{\sqrt2}\eta_{ab}\Phi^n_{\ nc}+\Phi_{abc}\right)\\
		&=\int d^3x\left(\frac{1}{4}\Phi_m^{\ mc}\Phi^n_{\ nc}-\frac{1}{6}\Phi^{abc}\Phi_{abc}\right)\\
		&=S_{LG}[\rho]\ ,\label{S(rho)}
		\end{split}
	\ee
which is LG in 3D. The sign $\pm$ in \eqref{rho} depends on the sign of $\kappa_m$ as a consequence of the contact \eqref{hc xi1=0} for which $\kappa_m=\pm2\sqrt{\kappa\tilde\kappa}$. Notice that if $\xi_1=0$, $S_{bd}$ \eqref{LG-Sbd} becomes
	\be\label{Sbdhc1}
	S_{bd}=\xi_0\int d^4x\delta(x^3)\left( h_{ab}h^{ab}-h^2\right)\ ,
	\ee
$i.e.$ the boundary action $S_{bd}$ \eqref{LG-Sbd} does not depend on the $\partial_3$-derivative of the gauge field anymore. We recognize in $S_{bd}$ the Fierz-Pauli mass term \cite{Hinterbichler:2011tt,Blasi:2017pkk,Blasi:2015lrg,Gambuti:2020onb,Gambuti:2021meo}, 
which renders the relation with LG even more remarkable. This allows to interpret $\xi_0$ as a Fierz-Pauli mass for the tensor field $h_{ab}(X)$ on the boundary $x^3=0$.
	\item $\pmb{\xi_1\neq 0}$ : the BC are the following
	\begin{empheq}{align}
	%\xi_1
	H^{ab}&=0\label{bcH}\\
	2\xi_0 \xcancel{H^{ab}}+(2\lambda+\xi_1)\tilde H^{ab}&=0\ .\label{bcH+tH}
	\end{empheq}
We have to distinguish between two cases: $2\lambda+\xi_1=0$ and $2\lambda+\xi_1\neq0$. For $\xi_1=-2\lambda$ we are left with the BC \eqref{bcH} only, which depends on $h_{ab}(X)$, hence on $\sigma_{ab}(X)$ through the solution \eqref{solh}. To have a contact, we have to switch off the $\tilde\sigma_{ab}(X)$ dependence in the 3D action \eqref{Sinv'} (and in the EoM \eqref{eomH1''}, \eqref{eomH2''}) by putting $\kappa_m=\tilde\kappa=0$. The induced theory in this case is LG for the field $\sigma_{ab}(X)$
	\be
	S_{3D}=\kappa S_{LG}=\kappa\int d^3x\left(\tfrac{1}{4}f^a_{\ ac} f_b^{\ bc}-\tfrac{1}{6}f^{abc}f_{abc}\right)\ ,
	\ee
where we can reabsorb the $\kappa$ parameter through a redefinition of the field $\sigma_{ab}(X)$. The boundary action term $S_{bd}$ \eqref{LG-Sbd} is 
	\be
	S_{bd}=\int d^4x\delta(x^3)\left[\xi_0 (h_{ab}h^{ab}-h^2)-2\lambda(\tilde h_{ab}h^{ab}-\tilde hh)\right]\ .
	\ee
Notice that also in this case the $\xi_0$ parameter plays the role of a Fierz-Pauli mass for $h_{ab}(X)$ on the boundary.
If instead $\xi_1\neq\{-2\lambda,0\}$ 	\begin{empheq}{align}
	H^{ab}&=0\label{BCH}\\
	\tilde H^{ab}&=0\ .\label{BCtH}
	\end{empheq}
Looking at the  3D boundary-side (EoM) we can use the EoM \eqref{eomH1}
	\be
	\tilde H^{ab}=-2\mu\frac{\kappa}{\kappa_m}H^{ab}\label{eomH1'}
	\ee
in the EoM \eqref{eomH2}, which becomes
	\be
	\frac{\mu}{2}\left(\frac{\kappa_m^2-4\kappa\tilde\kappa}{\tilde\kappa\kappa_m}\right)H^{ab}=0\quad;\quad\tilde\kappa,\ \kappa_m\neq0\ .\label{eomH2'}
	\ee
Now we notice that if $\kappa^2_m-4\kappa\tilde\kappa=0$ the EoM \eqref{eomH2'} becomes trivial, and we only have one EoM, which is \eqref{eomH1'}, which can never match the two BC \eqref{bcH} and \eqref{bcH+tH} at the same time. Indeed this case ($\kappa^2_m-4\kappa\tilde\kappa=0$) allows a contact only if we look at the BC in the form \eqref{bc2H} and \eqref{bc2''} and set $\xi_1=0$, which coincide with Case 1 \eqref{hc xi1=0}. Therefore $\kappa^2_m-4\kappa\tilde\kappa=0\ \Leftrightarrow\ \xi_1=0$. Considering $\kappa^2_m-4\kappa\tilde\kappa\neq0$ we can use the second EoM \eqref{eomH2'} back into the first one \eqref{eomH1'} and get
	\begin{empheq}{align}
	H^{ab}&=0\\
	\tilde H^{ab}&=0\quad;\quad\kappa^2_m-4\kappa\tilde\kappa\neq0,\ \tilde\kappa,\ \kappa_m\neq0\ ,
	\end{empheq}
which matches exactly the BC \eqref{BCH} and \eqref{BCtH}. Thus the holographic contact is possible for $\xi_1\neq\{-2\lambda,0\}$, $\kappa_m\neq\{0,2\sqrt{\kappa\tilde\kappa}\}$ and $\tilde\kappa\neq0$. Again $\xi_0$ does not affect the contact and can be interpreted as a Fierz-Pauli mass.
	\end{enumerate}
We summarize our results in the following Table \ref{summary}
\begin{table}[H]
	\begin{tabular}{|c|c|c|c|}
	\hline
	$\pmb{S_{bd}}$ {\bf parameters}&{\bf Constraints}&$\pmb{S_{bd}=}$&$\pmb{S_{3D}=}$\\\hline
	$\xi_1=0,\ \xi_0\mbox{ free}$&$\kappa=\frac{\xi_0}{2\mu\lambda}\kappa_m\ ;\ \tilde\kappa=\frac{\mu\lambda}{2\xi_0}\kappa_m\ ;\ \kappa_m^2=4\kappa\tilde\kappa$ &$S_{bd}[h]$&$S_{LG}[\rho]$\\\hline
	$\xi_1=-2\lambda,\ \xi_0\mbox{ free}$&$\kappa\ \mbox{free}\ ;\ \tilde\kappa=0\ ;\ \kappa_m=0$&$S_{bd}[h,\tilde h]$&$S_{LG}[\sigma]$\\\hline
	$\xi_1\neq\{-2\lambda,0\},\ \xi_0\mbox{ free}$&$\kappa\ \mbox{free}\ ;\ \tilde\kappa\neq0\ ;\ \kappa_m\neq\{0,2\sqrt{\kappa\tilde\kappa}\}$&$S_{bd}[h,\tilde h]$&$\kappa S_{LG}+\tilde\kappa\tilde S_{LG}+\kappa_mS_{mix}$\\\hline
	\end{tabular}
\caption{\footnotesize{Scheme of the contacts between bulk (BC) and boundary (EoM) with constraints on the parameters of $S_{bd}$ \eqref{LG-Sbd} and of $S_{3D}$ \eqref{Sinv'}.}}
\label{summary}
\end{table}
As we see from Table \ref{summary}, depending on the value of the $\xi_1$ parameter of $S_{bd}$ \eqref{LG-Sbd}, we found two possibilities for the 3D theory induced by the presence of a planar boundary on the 4D LG theory
	\bi
	\item $\pmb{\xi_1=\{0,-2\lambda\}}$ the 3D induced theory is LG for one symmetric rank-2 tensor field
		\be
		S_{3D}=S_{LG}\ .
		\label{S3D1}\ee
The corresponding 4D $S_{bd}$ \eqref{Sbdhc1} is a Fierz-Pauli mass term for the 4D tensor field $h_{ab}(X)$ whose mass parameter is $\xi_0$.
	\item $\pmb{\xi_1\neq\{0,-2\lambda\}}$ the 3D induced action depends on two rank-2 symmetric tensor fields $\sigma_{ab}(X)$ and $\tilde\sigma_{ab}(X)$. After a field redefinition, it reads
		\be
		S_{3D}[\sigma,\tilde\sigma]=S_{LG}[\sigma]+\tilde S_{LG}[\tilde\sigma]+kS_{mix}[\sigma,\tilde\sigma]\ ,
		\label{S3D2}\ee
where $k$ is a constant which cannot be reabsorbed.
	\ei
In both cases, the 4D boundary term $S_{bd}$ \eqref{LG-Sbd} contains a Fierz-Pauli mass term for the bulk tensor field $h_{ab}(X)$, whose mass parameter is $\xi_0$.

%*****************************************************
\section{Summary of results}
%*****************************************************

In this paper we studied the effect of the presence of a planar boundary on 4D Linearized Gravity (LG), realized by means of a Heaviside step function in the action \eqref{LG-Sbulk}. Following a method introduced by Symanzik in \cite{Symanzik:1981wd}, we derived the most general boundary conditions (BC) \eqref{BCh}, \eqref{bc2} and \eqref{bc3} by means of a variational principle. The boundary action $S_{bd}$ \eqref{LG-Sbd}, and hence the BC, depend on four parameters, which in our approach play an important role, as we shall comment on later. The presence of the boundary breaks the invariance under diffeomorphisms, which are the symmetry transformations of LG. Correspondingly, the Ward identity which describes the invariance under diffeomorphisms \eqref{LG-ward1} acquires a breaking, which is crucial, because from it the main information of the theory might be derived, namely the fields content, the symmetry transformations and the boundary algebra. We wrote ``might'' because it is not obvious that this can always be done. In fact, this seems to work for all topological field theories, where non trivial boundary dynamics has been first observed \cite{Amoretti:2014iza}, and for a long time this property has been believed to be peculiar of these kind of theories. More recently, similar results have been found in non topological field theories, like Maxwell theory \cite{Bertolini:2020hgr}, and this motivated boundary investigations for more general theories, like we did in this paper for LG. A first remarkable result is that on the boundary we found two conserved currents \eqref{K} and \eqref{Kt} which form the algebraic structure \eqref{KK1} and \eqref{KK2} of the Kac-Moody (KM) type, whose central charge is proportional to the inverse of the LG ``coupling'' constant \eqref{cc}. This confirms what has been guessed in \cite{Hinterbichler:2022agn}, where it was suspected the existence, in 4D LG, of a KM algebra as a particularly interesting possibility in connection with Weinberg's soft graviton theorems \cite{steve,He:2014laa,Kapec:2015vwa}.
Since the central charge of a KM algebra must be positive, this is mostly useful to determine the sign of the overall LG action, which otherwise should be determined by imposing that the energy density, that is the 00 - component of the energy-momentum tensor, is positive, which in gravity is a known tricky issue \cite{Carroll:2004st,Misner:1973prb}. Moreover, we were able to solve the on-shell Ward identity \eqref{cc1'} and the universal BC \eqref{cc2'} getting \eqref{solht} and \eqref{solh}, which allowed us to express, on the boundary, the 4D bulk fields $h^{ab}(X)$ and $\tilde h^{ab}(X)$ in terms of 3D fields which are the degrees of freedom of the induced 3D theory. We found that these latter, like their 4D ancestors, are rank-2 symmetric tensor fields: $\sigma^{ab}(X)$ and $\tilde \sigma^{ab}(X)$. This, as LG shows, seems to be peculiar of non topological QFTs. Indeed what is usually found in topological QFTs is that the fields living on the $D-1$-dimensional boundary are tensors of lower rank with respect their $D$-dimensional counterpart: from rank-2 tensors one finds vectors in the topological 4D BF theory \cite{Amoretti:2014iza} and the boundary reduction of the gauge field in Chern-Simons theory gives scalars. Here, instead, the 3D boundary fields are rank-2 symmetric tensor fields as those of 4D LG \cite{Blasi:2022mbl}. And, quite interestingly, the transformation which keeps invariant the definition of the boundary fields turns out to be the diffeomorphisms \eqref{tdiff} and \eqref{diff}, which therefore are a consequence of the introduction of the boundary, rather than an {\it a priori} request. Given the dynamical fields and the symmetry transformations, requiring locality and power counting allowed us to find the most general 3D action $S_{3D}$ \eqref{Sinv'}, which consists of three terms. Each term being invariant by its own, $S_{3D}$ depends on three constants which we do not reduced by redefining the 3D fields as we could, but we fixed them by establishing a ``holographic'' contact as our last step. This has been realized by requiring that the equations of motion of the 3D action $S_{3D}$ coincide with the BC of the 4D theory. To do that, we had at our disposal the 4 parameters on which $S_{bd}$ \eqref{LG-Sbd} depends and the three constants in $S_{3D}$ \eqref{Sinv'}. As an outcome of this tuning, we found two possibilities, depending on the value of one particular parameter appearing in $S_{bd}$~: $S_{3D}$ describes either LG for one single tensor field \eqref{S3D1}, or the action \eqref{S3D2}, containing two decoupled LG terms for the boundary tensor fields $\sigma^{ab}(X)$ and $\tilde \sigma^{ab}(X)$ and one term which mixes them. As a last, but probably not least, fact, we remark that in any case the $S_{bd}$ action term which governs the holographic contact contains a mass term \eqref{Sbdhc1} for the bulk tensor field $h_{ab}(x)$ of the particular Fierz-Pauli type \cite{Hinterbichler:2011tt,Blasi:2017pkk,Blasi:2015lrg,Gambuti:2020onb,Gambuti:2021meo}, with a free parameter $\xi_0$ which we can interpret as a mass.
    
 %********************************************************************
\section*{Acknowledgments}
%********************************************************************

We thank Alberto Blasi for enlightening discussions. This work has been partially supported by the INFN Scientific Initiative GSS: ``Gauge Theory, Strings and Supergravity''. E.B. is supported by MIUR grant ``Dipartimenti di Eccellenza'' (100020-2018-SD-DIP-ECC\_001). 

%********************************************************

\end{document}